\documentclass[aps,prb,superscriptaddress,twocolumn]{revtex4}
\usepackage{amsmath,epsfig,color,amssymb}
\usepackage{graphicx}
\usepackage{graphicx}
\usepackage{dcolumn}
\usepackage{bm}
\usepackage{graphicx}
\usepackage{amsmath, amsfonts, amssymb,mathrsfs}
\usepackage{pstricks}
\usepackage{amsxtra}
\usepackage{amsthm}
\usepackage{natbib}

\usepackage{color}

\definecolor{jrp}{rgb}{.05,.80,.8}

\definecolor{mjg}{rgb}{.08,.05,.8}

\definecolor{yyl}{rgb}{.8,.05,.08}

\newcommand{\delete}[1]{{}}

\def\be{\begin{equation}}
\def\ee{\end{equation}}
\def\bea{\begin{eqnarray}}
\def\eea{\end{eqnarray}}
\newcommand{\ket}[1]{\mbox{$|#1\rangle$}}
\newcommand{\bra}[1]{\mbox{$\langle#1|$}}

\def\be{\begin{equation}}      
\def\ee{\end{equation}}
\def\beu{\begin{equation*}}   
\def\eeu{\end{equation*}}

\providecommand{\abs}[1]{\left\lvert#1\right\rvert}   
\providecommand{\ket}[1]{\left|#1\right\rangle}

\providecommand{\bra}[1]{\left\langle#1\right|}

\providecommand{\del}{\partial}

\graphicspath{}

\begin{document}
\title{High-Order Multipole Radiation from Quantum Hall States in Dirac Materials}
\date{\today}
\author{Michael~J.~Gullans}
\author{Jacob M.~Taylor}
\affiliation{Joint Quantum Institute, NIST and University of Maryland, College Park, Maryland 20742, USA}
\affiliation{ Joint Center for Quantum Information and Computer Science, NIST and University of Maryland, College Park, Maryland 20742, USA.}
\author{Ata{\c c} Imamo{\u g}lu}
\affiliation{Institute of Quantum Electronics, ETH Z{\" u}rich, CH-8093 Z{\" u}rich, Switzerland.}
\author{Pouyan Ghaemi}
\affiliation{Physics Department, City College of the City University of New York, New York, NY 10031, USA.}
\affiliation{Physics Department, Grad. Center, City University of New York, New York, NY 10031, USA.}
\author{Mohammad Hafezi}
\affiliation{Joint Quantum Institute, NIST and University of Maryland, College Park, Maryland 20742, USA}
\affiliation{Department of Electrical and Computer Engineering, IREAP, University of Maryland, College Park, Maryland 20742, USA.}
\begin{abstract}
We investigate the optical response of strongly disordered quantum Hall states in two-dimensional Dirac materials and find qualitatively different effects in the radiation properties of the bulk versus the edge.
We show that the far-field radiation from the edge is characterized by large multipole moments ($> 50$) due to the efficient transfer of angular momentum from the electrons into the scattered light.  The maximum multipole transition moment is a direct measure of the coherence length of the edge states.  Accessing these multipole transitions would provide new tools for optical spectroscopy and control of   quantum Hall edge states.    On the other hand, the far-field radiation from the bulk appears as random dipole emission with spectral properties that vary with the local disorder potential.  We determine the conditions under which this bulk radiation can be used to image the disorder landscape.  Such optical measurements can probe sub-micron length scales over large areas and provide complementary information to scanning probe techniques.   Spatially resolving this bulk radiation would serve as a novel probe of the percolation transition near half-filling.
\end{abstract}
\maketitle

\section{Introduction}
The advent of graphene and other 2D materials has significantly increased the number of optically accessible, two-dimensional electron systems that exhibit the quantum Hall effect. \cite{Novoselov05,Zhang05,Fallahazad16,Li16}  These materials can be engineered into devices with nearly atomic scale precision, enabling advances in the manipulation  and spectroscopy of quantum Hall states \cite{Lee17,Wei17}.   As compared to low-frequency transport and electrical control, optical methods do not require Ohmic or superconducting contacts and can be reconfigured on sub-micron length and sub-ps time scales.  Motivated by the prospect for quantum optical manipulation of quantum Hall states in these materials, we investigate fundamental effects in their optical response when the wavelength of light is much less than the size of the sample.  This knowledge can be used to design optical-based protocols for spatially resolved manipulation and spectroscopy of quantum Hall states.

Optical studies of quantum Hall systems display a rich phenomenology due to  the strong effect the magnetic field has upon the electronic orbitals and levels.  
For laboratory magnetic fields, intra-band Landau level transitions typically lie in the far-infrared (IR) portion of the electromagnetic spectrum \cite{McCombe75,Kono11,Maag16,Curtis16,Dmitriev12}.
The long wavelength of these transitions enables  several novel applications to quantum optics \cite{Aoki86,Morimoto08,Mittendorff14,Wendler15,Wang15,Hagenmuller12,Chirolli12,Scalari12,Pellegrino14,Kono16,Zhang16}, but increases experimental difficulty.  
Interband transitions can cover a wide range of wavelengths depending on the band structure and have been extensively studied in AlGaAs heterostructures for spectroscopy of fractional quantum Hall states \cite{Heiman88,Turberfield90,Goldberg90,Buhmann90,Hawrylak03,Byszewski06,Smolka14}. 
 Inter-Landau level transitions in graphene have been spectroscopically probed from terahertz up to optical frequencies 
\cite{Jiang07,Plochocka09,Orlita11,Maero14,Funk15,Faugeras11,Goler12,Faugeras15,Nazin10}.  
In the transition metal dichalcogenides, the magneto-optical response is typically dominated by excitonic effects due to the large exciton binding energy in these materials \cite{Li14,Srivastava15,MacNeil15,Urbaszek15,Aivazian15,Mitioglu15,ChuR14}.
However, optical signatures of inter-band Landau level transitions have been directly observed in WSe${_2}$ \cite{Wang16}.



In this article, we investigate 2D materials whose low-energy band structure can be approximately described by a Dirac model, which we refer to as 2D Dirac materials (2DDM).  We show for the first time that the quantum Hall edge states support high-order, radiative multipole transitions.  These transitions are a consequence of the large electronic coherence length and topological translation symmetry of the edge states, but have been overlooked in previous treatments of the optical response of quantum Hall systems.    Accessing these transitions would allow novel methods for optical spectroscopy and manipulation of integer and, potentially, fractional  quantum Hall edge states.  
On the other hand, the radiation from the bulk of the 2DDM is dominated by dipole emission, whose spectral properties are correlated with the disorder landscape.   We find the conditions under which these bulk optical transitions can be spatially resolved, which enables optical imaging and manipulation of the potential landscape of the quantum Hall states.

Consider a 2DDM  in the integer quantum Hall regime with an electron-hole pair excited above the Fermi level.  At integer filling, standard arguments show that the majority of the states in the bulk are localized due to disorder \cite{HalperinLaughlin8182}.  
 When the localization length of the electron-hole pair is much less than the optical wavelength, the optical radiation in the far field will appear as dipole emission, but with a spectrum that varies with the local disorder potential [see Fig.~1(a)].
This arguments demonstrates that spatially mapping out the emission spectrum across the sample will reveal correlations in the disorder on the scale of the optical wavelength.

As the electron-hole pair approaches the edge, the situation changes dramatically because these states are not localized and exhibit electronic coherence that extends across the entire sample \cite{HalperinLaughlin8182}.  Furthermore, due to the magnetic field, the edge states carry a large  angular momentum.  In principle, this angular momentum can be transferred into the optical radiation during emission.  Such a transfer process is necessarily associated with the presence of higher order multipole moments in the far-field radiation.

To examine the nature of the spontaneously emitted radiation, we also decompose the optical field into eigenmodes of $L_z$ about the center of the 2DDM sample with orbital angular momentum (OAM) $\hbar \ell$ and longitudinal momentum $\hbar k$.  Such states are known as cylindrical vector harmonics  and are closely related to the cylindrically symmetric Laguerre-Gaussian modes within the paraxial approximation \cite{Craig83}.
Due to disorder, the electrons on the edge will not be in a pure angular momentum eigenstate, but will be in a superposition of angular momentum states narrowly peaked around the value $m_e \sim r_e^2/\ell_c^2$, where $r_e$ is the approximate radius of the edge and $m_e$ is the angular momentum quantum number defined in a gauge-invariant manner in Appendix \ref{sec:appA}.  The multipole transitions arise because any  electron in the conduction band in the angular momentum state $ m$ can conserve total angular momentum by recombining with a  hole in the valence band in the state $ m'$ and  emitting light with OAM $\ell  = m- m'$ [see Fig.~1(b)].  We find that these transitions are allowed with a nearly uniform branching ratio up to a cutoff give by $2\pi r_e/\lambda$, where $\lambda$ is the optical wavelength.  When the dephasing of the electron transport on the edge is included, this scaling should be modified to $\ell_\phi/\lambda$, where $\ell_\phi$ is the coherence length of the  edge states. 

These arguments are quite general and demonstrate that the multipole radiation is a direct consequence of the large electronic coherence length of the edge states.  To understand the behavior and scaling of these transitions in more detail, we consider a cylindrically symmetric edge below such that the multipole radiation pattern can be calculated analytically.  

  \begin{figure}[t]
\begin{center}
\includegraphics[width=0.45 \textwidth]{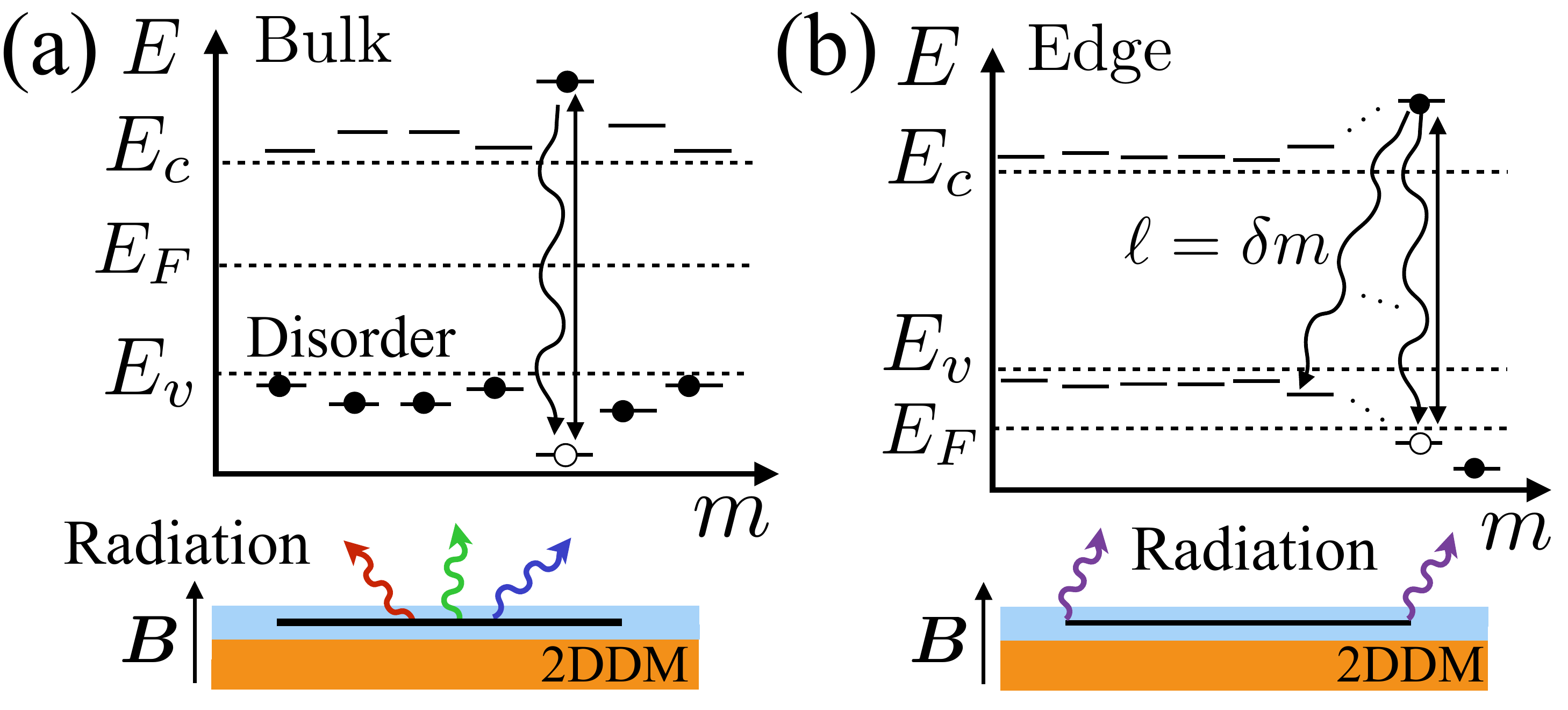}
\caption{(a) In the presence of a large magnetic field, the electronic states of the 2DDM are quantized into Landau levels, which we index by their angular momentum $-\hbar m$.  The majority of the states in the bulk are localized by disorder, leading to inter-band  radiation dominated by dipole emission.  The spectrum of this radiation is spatially correlated with the  disorder potential.  Here $E_{c (v)}$ refer to the energy of the conduction (valence) band and $E_F$ is the Fermi energy.  (b) An electron  excited at the edge of the system can emit light with orbital angular momentum $\hbar \ell$ by recombining with a hole in the state $m'=m-\ell$.  
}
\label{default}
\end{center}
\end{figure}

%

\section{Dirac Model}

We consider the low-energy Dirac Hamiltonian of the forrm (neglecting spin),
\be
H= \hbar v \,\bm{k} \cdot \bm{\tau} + m_0 v^2 \tau_z ,
\ee
where $v$ is the Dirac velocity, $\bm{k}=(k_x,k_y)$ is the in-plane wavevector, $\bm{\tau}=(\tau_x,\tau_y,\tau_z)$ are Pauli matrices operating on the Dirac pseudospin, and $m_0$ is the effective Dirac mass.  At zero magnetic field the spectrum of $H$ is $\epsilon(\bm{k})=\pm \sqrt{m_0^2 v^4 +v^2 |\bm{k}|^2}$. For large $B_z$, the energy spectrum is quantized into degenerate Landau levels at energies $\epsilon_n= {\rm sign}(n)\sqrt{m_0^2 v^4 + \hbar^2 \omega_c^2 |n|}$, where $n$ is an integer, $\omega_c=\sqrt{2} v/\ell_c$ is the cycolotron frequency, and $\ell_c =\sqrt{\hbar /eB_z}$ is the magnetic length.   Throughout this work we restrict our discussion to a single valley for simplicity.

  The light-matter interaction for $H$ can be found  through the usual prescription $\bm{k} \to \bm{k} - e \bm{A} /c$
  \be \label{eqn:hint}
 H_{\rm int}=  \frac{e v}{\sqrt{2} c} [\tau_+ A_+^*(x,y) + \tau_- A_-^*(x,y)]e^{-i \omega t} + h.c.,
 \ee
  where $A_\pm=(A_x\pm i A_y)/\sqrt{2}$ are the $\sigma_\pm$ circularly polarized components of the vector potential $\bm{A}$ in the plane of the 2D material.   Due to the Dirac band structure, the pseudo-spin operators $\tau_\pm$ couple the $n$th Landau level to both $n \pm 1$ and $-n \pm 1$.  This leads to the optical selection rule for $\sigma_\pm$ circularly polarized light: $n \to n'$ with $\abs{n'}=\abs{n} \pm 1$. \cite{Jiang07}

  We represent the single-particle states in the symmetric gauge, in which case the eigenstates $\ket{n,m}$ take the form \cite{WenBook}
    \be \label{eqn:psiz}
\begin{split}
\bra{x,y}&{n,m}\rangle \propto 
 \left( \begin{array}{c}
\alpha_n \sqrt{|n|} D_{\bar{u}}^{|n|-1} \bar{u}^{|n|+m}  \\
\beta_n  \sqrt{2} i \ell_c D_{\bar{u}}^{|n|} \bar{u}^{|n|+m} 
\end{array} \right) e^{-|u|^2/4 \ell_c^2},
\end{split}
\ee
where $u=x+iy$,  $D_{\bar{u}}=  \del_{\bar{u}} - u/2\ell_c^2$ acts as a raising operator on the Landau level eigenfunctions,  $(\alpha_0, \beta_0)^T = (0, 1)$, and, for $n>0$~$(n<0)$, $(\alpha_n, \beta_n)^T$  are the positive (negative) eigenvectors of the 2x2 matrix 
\be \label{eqn:hn}
H_n=\left(\begin{array}{c c}
m_0 v^2 & \hbar \omega_c \sqrt{|n|} \\
\hbar \omega_c \sqrt{|n|} &- m_0 v^2
\end{array} \right),
\ee
whose eigenvalues are the energy eigenvalues $\epsilon_n$.
  We represent the OAM eigenstates for the optical field in the basis of cylindrical vector harmonics \cite{Craig83}, which take the form $\bm{E}(x,y,z) = \sum_{\ell,k} \bm{E}_{\ell,k}(r) e^{i \ell \theta+i kz}$, where $r=|u|$ and $\theta = \tan^{-1}(y/x)$.

\section{Radiation from the Edge}

We first consider the light emission from the edge states of the quantum Hall system.    The edge can  either be formed by an external confining potential, at an interface with vacuum or another material, or from an abrupt change in the local dielectric environment.    
An externally applied  potential will generally lead to identical confining potentials for the Landau levels in the conduction and valence band.  As a result, the optical transitions between edge states will be degenerate with the transitions in the bulk.  

In order to selectively address the edge states, it is desirable to a have a difference in dispersion between the edge states in the conduction and valence bands [see Fig.~1(b)].  Such a difference in slope can arise at a sharp interface due to local modifications of the band structure \cite{Abanin07}.  In the case of graphene with a vacuum interface, the dispersion of the quantum Hall edge states depends on whether the edge termination is of armchair or zig-zag type \cite{Abanin07}.   For $|n|>0$, however, all edge states disperse with the opposite sign in the conduction and valence band, which allows these optical transitions to be spectrally distinguished from the bulk.   This analysis can be generalized to include a Dirac mass and one finds that the opposite slope of the conduction and valence band is preserved. 
Alternatively, 
one can consider an edge formed by a change in the local dielectric environment, e.g., an additional layer of insulating material such as h-BN.  In this case, the change in the dielectric screening will modify the contribution of electron-electron interactions to the inter-band Landau level transitions \cite{Faugeras15}, which will result in  optically addressable edge states. 



  \begin{figure}[t]
\begin{center}
\includegraphics[width=0.45 \textwidth]{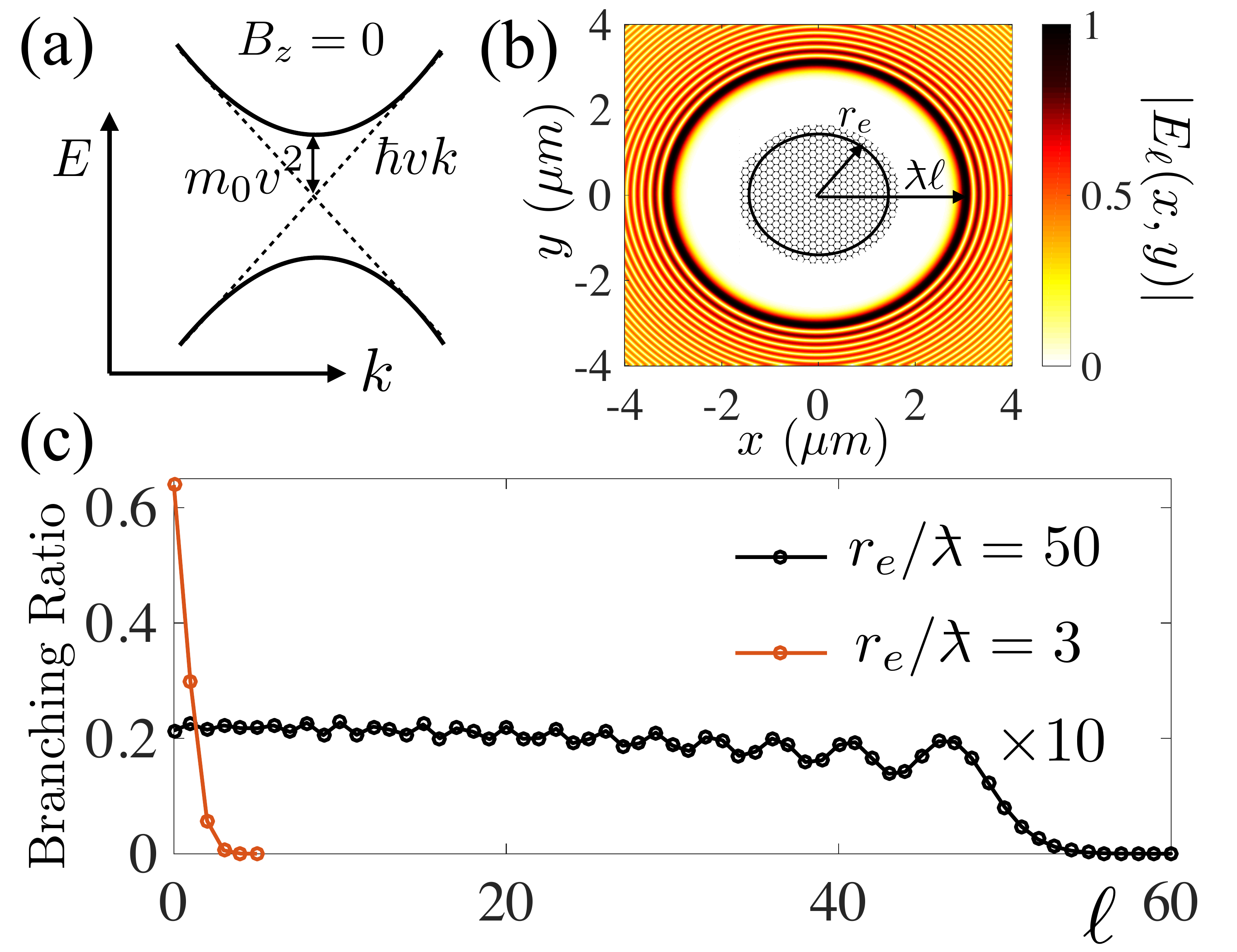}
\caption{(a) Low-energy band structure of graphene-like Dirac material for zero magnetic field.  Here $m_0$ and $v$ are the Dirac mass and velocity, respectively, and we only show one of the two valleys. (b) Amplitude of the cylindrical vector harmonic $|E_\ell |$ for $\ell = 100$ with $\lambda_0=600~$nm and index of refraction $n_0=3.2$.  Because the size of the optical vortex increases as $\lambdabar \ell$, an edge state with radius $r_e$ (black circle) can only spontaneously emit into modes with $\ell \lesssim r_e/\lambdabar$.
(c) Branching ratio for spontaneous emission into different $\ell$ modes for two different values of $r_e/\lambdabar$.  We took Dirac parameters for WSe$_2$ ($m_0 v^2 \approx 1$~eV and $v \approx 10^6$~m/s \cite{Xiao12}) embedded in GaP, $B_z=11$~T,  $n=0$, and $\lambdabar = 30$~nm. }
\label{default}
\end{center}
\end{figure}


For the case of a cylindrically symmetric edge, the edge states are approximately given by the angular momentum states $\ket{n,m}$ whose size $r_m \approx \sqrt{m} \ell_c$ is equal to the radius of the edge  $r_e$.  As we noted above, 
one can achieve optical Raman transitions between edge states by transferring orbital angular momentum into the light field.
To understand the scaling of the multipole emission with increasing $\ell$, we note that light with OAM $\ell$ has an optical vortex in the center of size greater than or equal to $\lambdabar \ell$, where  $\lambdabar= \lambda/2\pi$ [see Fig.~2(b)].  Beyond this radius, the average intensity of the light is independent of $\ell$.  This implies that the emitted light will contain multipole contributions up to the maximum value $\ell_{\rm max}= r_e / \lambdabar$, where $r_e$ is the radius of the edge.  In addition, $\ell_{\rm max}$ will be cut off by the finite coherence length of the edge states $\ell_\phi$, arising from  electron-electron interactions, inter-valley scattering, and phonon scattering.  For integer quantum Hall states in GaAs, the coherence length was measured via transport methods to be at least 10~$\mu$m - 20~$\mu$m \cite{Roulleau08}, which is much greater than the relevant optical wavelengths.  


To understand this effect more quantitatively, we decompose the radiative emission rate $\gamma_m$ of an excited electron in the state $\ket{n+1,m}$ into all the multipole moments $\gamma_m = \sum_{\ell\ge0} \gamma_m^\ell$ \cite{footnote_ell}.  Each individual component can be found using Fermi's golden rule for the emission into the free space modes with a specified $\ell$.  We give the matrix elements in Appendix B.  Two illustrative examples are shown in Fig.~2(c) for the $n=0$ to $n=1$ transition with Dirac parameters for single-layer WSe$_2$.  We plot the branching ratio $\gamma_m^\ell/\gamma_m$ for two different edge radii, which confirms the scaling analysis from above.  For $r_e = 1.5~\mu$m we find a nearly uniform distribution for the spontaneous emission out to $\ell = 50$.  
Including disorder will modify shape of the distributions in Fig.~2(c), but it will not reduce $\ell_{\rm max}$, which is simply a result of the large coherence length of the edge states compared to $\lambdabar$.

\section{Radiation from the Bulk}

We now consider the optical emission from the localized states in the bulk of the 2D material at integer filling.  In particular, we show that the disorder landscape can be reconstructed through optical imaging of the scattered light.
%
We can include disorder in the Dirac model by adding all terms consistent with the symmetries of the hexagonal lattice (neglecting inter-valley scattering) \cite{DasSarma11}
\be
H_{\rm dis} = u_0(\bm{r}) I + \bm{u}(\bm{r}) \cdot \bm{\tau}.
\ee
The first term $u_0$ corresponds to long range diagonal disorder arising from, e.g., charged impurities, while the other terms are associated with shorter range effects such as, e.g., variations in the two sub-lattice potentials ($u_z$), tunneling rates $(u_{x,y})$, or the presence of vacancies and defects.

The projection of $H_{\rm dis}$ into the Landau levels leads to smoothing of the  disorder on the scale of $\ell_c$. This produces a potential landscape for each Landau level $U_n(x,y)= \bra{x,y}{\rm Tr}_\tau (P_{n} H_{\rm dis} P_{n}) \ket{x,y}$, where  $P_{n}$ is a projector into the $n$th Landau level and ${\rm Tr}_\tau$ traces over the pseudospin states. This landscape gives rise to (1) an adiabatic shift of the edge position and (2)  localized states in the bulk. Thus, the edge multipole effects remain the same, while the bulk radiation becomes dominated by transitions between localized states, each with a different spectral signature [see Fig.~1(a)].  

   \begin{figure}[t] 
\begin{center}
\includegraphics[width=0.48 \textwidth]{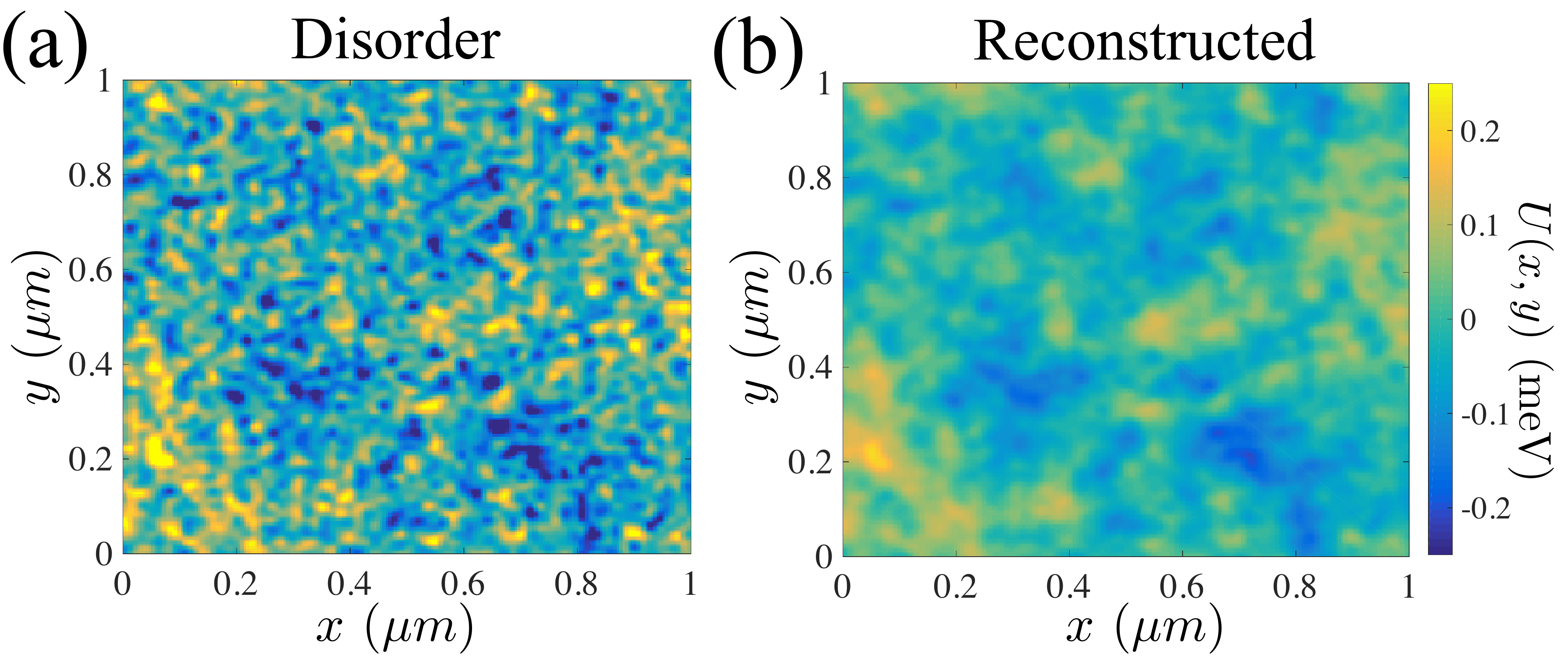}
\caption{(a) The disorder potential $U(x,y)$ for the inter-band  transitions between Landau levels.   (b) $U(x,y)$ can be reconstructed  by correlating the amplitude of spatially-resolved scattered light with the  frequency of the incoming probe.    We took the 2DDM to be embedded in GaP $(n_0=3.2)$ in a 10~T magnetic field with $\lambda_0=1~\mu$m.  The optical imaging is able to resolve spatial features down to the diffraction limit $\lambda_0/2 n_0 \approx 160$~nm. 
}
\label{default}
\end{center}
\end{figure}

To see how these spectral signatures can be used to image the disorder landscape,
we consider near resonant excitation between Landau levels  with $\sigma_+$ polarized light and a probe whose frequency $\omega_\ell$ is scanned through the resonance $\hbar \omega_\ell = \epsilon_{n+1}-\epsilon_{-n}$.  The disorder in the optical transition frequency $U(x,y)=U_{n+1}(x,y)-U_{-n}(x,y)$ for $n=0$ is shown in Fig.~3(a).  
To obtain the spatial profile of emitted light we approximate the far field emission pattern by a convolution of $U(x,y)$ with the filter function  $\eta_{\lambda}(r) = \sin (4 \pi r/\lambda )/ \pi^2 r^2$, which arises from the diffraction limit.  Here  $\lambda= [(h n_0/c) (\epsilon_{n+1}-\epsilon_{-n})]^{-1}$ is the central wavelength of emitted light, and $n_0$ is the index of refraction of the surrounding substrate.  We construct the disorder potential by finding the probe frequency at which the local scattered light reaches its maximum amplitude.  The resulting optically reconstructed disorder potential is shown in Fig.~3(b).   In practice, this reconstruction will be limited by the numerical aperture NA of the imaging system.
The diffraction limit in free-space is NA$ \le 1$, using, e.g., a solid-immersion-lens, one can enhance the upper limit of the NA by the index of refraction of the lens
\cite{Wu99}.  Alternatively, super-resolution techniques would enable imaging far below the diffraction limit \cite{SubDiffraction,Tsang16}.  

As we are treating the disorder in degenerate, first-order perturbation theory, we can see from Eq.~(\ref{eqn:hn}) that, for massless Dirac Fermions,  $U(r)$ is  dominated by the $\tau_x$ disorder, while, for sufficiently massive Dirac fermions, $U(r)$ is dominated by  $\tau_z$ disorder.  A related measurement in massive 2DDMs could be used to indirectly map out the diagonal disorder term $u_0(r)$ by going away from integer filling.  In particular, the exciton binding energy will vary with the local carrier density due to screening effects.  Thus, mapping out the exciton line across the sample would reveal  variations in the local carrier density, which, in the partially filled, disordered quantum Hall regime, are directly correlated with the underlying disorder potential \cite{Efros93,Ilani04}.

\section{Electron-Electron Interactions}

In our analysis, we have largely neglected the effect of electron-electron interactions 
on both the disorder landscape and 
 the optically excited electron-hole pair.  
Near integer filling, the interactions will have a minimal effect on the bare disorder potential because the electronic state is incompressible and cannot screen the disorder \cite{Efros93,Ilani04}.  

The dominant effects of the electron-hole interactions is to lead to Landau level mixing and magnetexciton formation, which have to be considered separately for the bulk and the edge.  On the edge, magnetoxciton effects are weak because of the predominantly linear dispersion of the edge states. Landau level mixing can then also be ignored because the electron and hole are both  delocalized and interact weakly.  
For the bulk, our analysis assumes that the magnetoexciton binding energy $\epsilon_b$ is much less than the strength of the disorder potential. 
However, in the opposite limit of strongly bound excitons, the $\bm{\tau}$ disorder will lead to spatial variations in $\epsilon_b$.  As a result, we expect our conclusions about mapping the $\bm{\tau}$ disorder to remain valid in this limit, provided that the disorder potential contains long-range correlations compared to the magnetoexciton Bohr radius.

\section{Conclusion}

We have studied the properties of the optical radiation from integer quantum Hall edge states in Dirac materials.  
We showed that the optical emission from the bulk of the 2DDM reflects the disorder landscape and, at the edge, high-order multipole transitions become allowed, which have not been previously considered in the optical response of quantum Hall states.
As a result, this work establishes that high-order multipole radiation is an important component of the optical spectroscopy and control of quantum Hall states and related topological systems.  Furthermore, these large multipole moments may be useful for applications that make use of light with large orbital angular momentum \cite{Molina07}.  
Although in this work we have focused on effects which are independent of electron-electron interactions, extending the optical spectroscopy and control techniques described here to study  fractional quantum Hall systems or  magnetoexcitons is a rich avenue for further investigation.  

\begin{acknowledgements}
\emph{Acknowledgements.---}
We thank  A. MacDonald, G. Solomon, O. Gazzano, Y. Hatsugai, J. Shabani  and W. DeGottardi for helpful discussions. This research was supported in part by the Kavli Institute for Theoretical Physics through the NSF under Grant No. NSF PHY11-25915, NSF Grant No. EFRI-1542863, AFOSR-MURI FA95501610323, Sloan Fellowship, YIP-ONR, the NSF PFC at the JQI, and the NSF MRSEC at Princeton.
\end{acknowledgements}

\appendix
 \section{Gauge-Independent Derivation of Optical Selection Rules}
  \label{sec:appA}
  
  The Dirac Hamiltonian in the presence of a constant magnetic field in the $z$-direction can be diagonalized in a gauge independent manner by introducing the canonical momentum operators and guiding center coordinate operators ($\hbar=1$)
\begin{align}
\bm{\pi} &= \bm{k} + \frac{e \bm{A}_0}{c}, \\
\bm{R} &= (X,Y)=(x + \ell_c^2 \pi_y, y - \ell_c^2 \pi_x).
\end{align}
These operators satisfy canonical commutation relations $[\pi_x,\pi_y]= i /\ell_c^2$ and $[X,Y]=-i \ell_c^2$, which allows one to define commuting bosonic operators associated with these coordinates
\begin{align}
a &= \frac{i \ell_c}{\sqrt{2}} (\pi_x + i \pi_y),\\
b &= \frac{X - i Y}{\sqrt{2} \ell_c}.
\end{align}

In terms of these operators, the Hamiltonian takes the form
\be
H = i \omega_c (a^\dagger \tau_+ - a \tau_-) + m_0v^2 \tau_z,
\ee
which is independent of $b$.  We define the generalized angular momentum operator \cite{Kang16}
\be
L_z = a^\dagger a - b^\dagger b  - \tau_z/2 +1/2,
\ee
which commutes with $H$. In the symmetric gauge, $L_z = x k_y - y k_x - \tau_z/2+1/2$ is equivalent to the usual angular momentum operator with the added term $(1-\tau_z)/2$.  The simultaneous eigenstates of $H$ and $L_z$ in the K-valley are defined, for $n\ne 0$, as
\be
\ket{n,m} = \frac{(a^\dagger)^{\abs{n}-1} (b^\dagger)^{m+
 |n|-1}}{\sqrt{(m+|n|)!} \sqrt{|n|!}} \left(\begin{array}{c}
\alpha_n {\sqrt{(m+|n|)|n|}} \\
\beta_n  {a^\dagger b^\dagger}
  \end{array} \right)
  \ket{0}
 \ee
 and, for $n=0$, as
\be
\ket{0,m} = \frac{ (b^\dagger)^{m}}{\sqrt{m!} } \left(\begin{array}{c}
 0 \\
 1
  \end{array} \right)
  \ket{0}.
 \ee

 To understand the selection rules we consider a plane wave incident on the 2DDM with in plane circular polarization $\sigma_+$ and in-plane wavevector $k_\perp \hat{x}$ directed along the $x$-axis.  Using the representation for the position operator ${x} = \ell_c (b+b^\dagger + a+ a^\dagger)/\sqrt{2}$ we can write the light-matter interaction in a frame rotating with the optical field in terms of the quantum Hall creation and annihilation operators 
 \be
 H_\textrm{int} = A_0 ( \tau_+ e^{ - i k_\perp \ell_c (b+b^\dagger + a + a^\dagger)/\sqrt{2}} + h.c.).
 \ee
In this representation, we can see that the plane wave acts as a product of coherent state displacement operators $D_{a}(\alpha)D_b(\alpha)$ with amplitude $\alpha = i k_\perp \ell_c/\sqrt{2}$, i.e.,
\be
a\, e^{-i q \ell_c (a+a^\dagger)/\sqrt{2}} \ket{0} = aD_a(\alpha) \ket{0}=\alpha \ket{\alpha}.
\ee

Focusing on the $n=0$ state for simplicity, we see that acting with $H_\textrm{int}$ on $\ket{0,m}$ leads to the state
\be
\begin{split} \label{eqn:sel}
H_\textrm{int} \ket{0,m} &= A_0 D_b(\alpha) \frac{ (b^\dagger)^{m}}{\sqrt{m!} } \left(\begin{array}{c}
 D_a(\alpha) \\
 0
 \end{array} \right)
\ket{0} \\
&= A_0 \frac{ (b^\dagger-\alpha^*)^{m}}{\sqrt{m!} } \left(\begin{array}{c}
 D_a(\alpha) D_b(\alpha) \\
 0
 \end{array} \right)
\ket{0} 
\end{split}
\ee
To evaluate the selection rules we first note that we can neglect the effect of the  displacement operator $D_{a}(\alpha)$ in the second line of Eq.~(\ref{eqn:sel}) because $|\alpha| < \sqrt{2} \ell_c/\lambdabar \ll 1$ (here the first inequality follows because $k_\perp  < 2\pi/\lambda$).  Surprisingly, however, one is not justified in neglecting $\alpha$ in either the prefactor of this expression or in $D_b(\alpha)$.  To understand this result we expand Eq.~(\ref{eqn:sel}) into the basis $\ket{1,m}$ as
\begin{align} \nonumber
H_\textrm{int} \ket{0,m} &\approx A_0  \sum_{j=0}^m \binom{m}{j} \frac{(b^\dagger)^{m-j}(- \alpha^*)^{j}}{\sqrt{m!}} \left(\begin{array}{c}
 1 \\
 0
 \end{array} \right) \ket{0,\alpha}\\ 
 & = A_0 \alpha_1e^{-|\alpha|^2/2} \sum_{\ell} F_{m,\ell}(\alpha) \ket{1,m+\ell},\\ \label{eqn:fm}
 F_{m,\ell}(\alpha)&= \sqrt{\frac{(m+\ell)!}{m!}} \alpha^\ell \sum_{j=j_\ell}^m \binom{m}{j} \frac{(-1)^j |\alpha|^{2j}}{(\ell+j)!} ,
\end{align}
where $j_\ell= \max(0, - \ell)$.
Evaluating this sum and using Sterlings formula $n! \approx \sqrt{2 \pi n} (n/e)^n$, we find that the multipole moments are actually perturbative in $r_m k_\perp/\ell = \sqrt{m}\ell_c k_\perp/ \ell$ and not $\ell_c k_\perp/\ell$ as one would naively expect.  In particular, in the regime where  $r_m k_\perp/\ell<1$ we find the scaling
\be
\bra{1,m+\ell}H_\textrm{int} \ket{0,m} \sim  \bigg(\frac{ r_m k_\perp}{\ell}\bigg)^\ell,
\ee
which is identical to the scaling we find for the cylindrical vector harmonics in this regime.

For $r_m k_\perp/\ell > 1$, one has to use the nonperturbative expression from Eq.~(\ref{eqn:fm}) to evaluate the multipole transition moments.  Similar to the multipole radiation we found for the cylindrical vector harmonics, one finds (after averaging over $k_\perp$) that this expression is approximately independent of $\ell$ in this regime.
Thus we see that the gauge-independent representation of the plane wave  response is nearly identical to the response we found for the cylindrical vector harmonics discussed in the main text.

\section{Spontaneous Emission of Edge State in Symmetric Gauge}\label{sec:cvh}
In this section, we define the cylindrical vector harmonic solutions to Maxwell's equations.  We quantize these modes, give the expressions for the matrix elements used to calculate the spontaneous emission of the edge states, and evaluate the scaling of the spontaneous emission rate with increasing OAM.

To construct the cylindrical vector harmonics we start with  the cylindrically symmetric solutions to the Hemholtz equation
\be
\big(\nabla^2+k_0^2) \psi_{\ell,k}(\bm{r}) =0,
\ee
which take the form
\be
\psi_{\ell,k}(r,\theta,z) = e^{i k z +i \ell \theta}J_\ell(k_\perp r).
\ee
Here $(r,\theta,z)$ are the cylindrical coordinates such that $(x,y,z)=(r\cos \theta,r\sin\theta,z)$, $\ell$ is an integer that labels the orbital angular momentum,  $k$ is the longitudinal wavevector, $k_\perp = \sqrt{k_0^2-k^2}$, and $J_\ell(\cdot)$ are the Bessel functions of the first kind.  We can construct vector solutions as  \cite{Craig83}
\begin{align}
\bm{M}_{\ell,k} = \frac{\nabla \times (\hat{z}\, \psi_{\ell,k})}{k_\perp},\\
\bm{N}_{\ell,k} = \frac{\nabla \times \bm{M}_{\ell,k}}{k_0},
\end{align}
We can use these solutions to construct a complete basis for the the transverse solutions to Maxwell's equations in free space in terms of the vector potential in the Coulomb gauge
 \begin{align}
\bm{A}^1_{\ell,k} = A_0 \bm{M}_{\ell,k},  \\
\bm{A}^2_{\ell,k}= A_0  \bm{N}_{\ell,k},
\end{align}
where $A_0$ is the amplitude.  The energy density of $\bm{A}^i_{\ell,k}$ is given by 
\be
u = \frac{\omega^2 \epsilon_0}{2 k_\perp^2} (|\bm{M}_{\ell,k}|^2 + | \bm{N}_{\ell,k}|^2) |A_0|^2
\ee

We quantize these modes by placing them in a large cylindrical box of radius $R$ and length $L$.  After quantization the normalization constant $A_0$ is set by the condition $\int d^3 r\, u = \hbar \omega$, where $\omega = c k_0$
\be
A_0= \sqrt{\frac{\hbar k_\perp}{2 \epsilon_0 L R \omega}}
\ee

The key quantities that enter the calculations of the in the main text are the dipole matrix elements between the different Landau level states.  We now give explicit expressions for the matrix elements between the $n=0$ and $n=1$ Landau levels.
The $n=0$ and $n=1$ Landau level in the K-valley takes the form
\begin{align}
\ket{0,m}& = N_m^0 \left(\begin{array}{c}
0 \\
\bar{u}^m
\end{array}
\right) e^{-|u|^2/2 \ell_c^2},\\
\ket{1,m}& = N_m^1 \left(\begin{array}{c}
\alpha_1  \bar{u}^{m+1} \\
\beta_1 \sqrt{2} \ell_c i   \Big[m+ \frac{|u|^2}{2 \ell_c^2}\Big]  \bar{u}^{m} 
\end{array}
\right) e^{-|u|^2/2 \ell_c^2},
\end{align}
where 
\begin{align}
N_{m}^0&=\frac{\sqrt{2}}{\ell_c^{m}\sqrt{m !}},\\
 N_{m}^1&= \frac{\sqrt{2}}{\ell_c^{m+1}\sqrt{(m+1) !}} \frac{ i \sqrt{2} \ell_c}{\sqrt{|\alpha_1|^2 + |\beta_1|^2 (10+9 m)}}
 \end{align}
 are normalization constants.  For $n\le 0$ 
\begin{align}
\left(\begin{array}{c} \alpha_n \\ \beta_n \end{array}\right) &=  \frac{1}{\sqrt{ 2 |E_n|(|E_n|+ m_0v^2)}}\left(\begin{array}{c} \hbar \omega_c \sqrt{|n|} \\  m_0v^2+|E_n|  \end{array}\right),
\end{align}
and for $n>0$
\begin{align}
\left(\begin{array}{c} \alpha_n \\ \beta_n \end{array}\right) &=  \frac{1}{\sqrt{2 E_n(E_n+ m_0v^2)}}\left(\begin{array}{c}m_0v^2+E_n \\ - \hbar \omega_c \sqrt{n}   \end{array}\right),
\end{align}
where $E_n = \textrm{sign}(n) \sqrt{m_0^2v^4+\hbar^2 \omega_c^2 |n|}$.
The dipole matrix elements are given by
\be \label{eqn:m}
\begin{split}
M_{m',m}^{\ell,k,i}=&\bra{1,m'}\frac{e v}{\sqrt{2}} \tau_+ \bm{A}_{\ell,k}^i \cdot \hat{\sigma}_+^* \ket{0,m}\\
&=\frac{e v}{\sqrt{2}} \alpha_1 N_{m'}^1 N_{m}^0 \int dr\, r^{m'+m+1}e^{-r^2/\ell_c^2} \\
&\times \bm{A}_{\ell,k}^i \cdot\hat{\sigma}_+^* \delta_{m',m-\ell },
\end{split}
\ee
 $\hat{\sigma}_\pm = (\hat{x}\pm i \hat{y})/\sqrt{2} = e^{\pm i \theta} (\hat{r} \pm i \hat{\theta})/\sqrt{2}$ and $\delta_{nn'}$ is the Kronecker delta function.  These integrals can be expressed analytically in terms of hypergeometric functions.   


\begin{figure}[t]
\begin{center}
\includegraphics[width=.45 \textwidth]{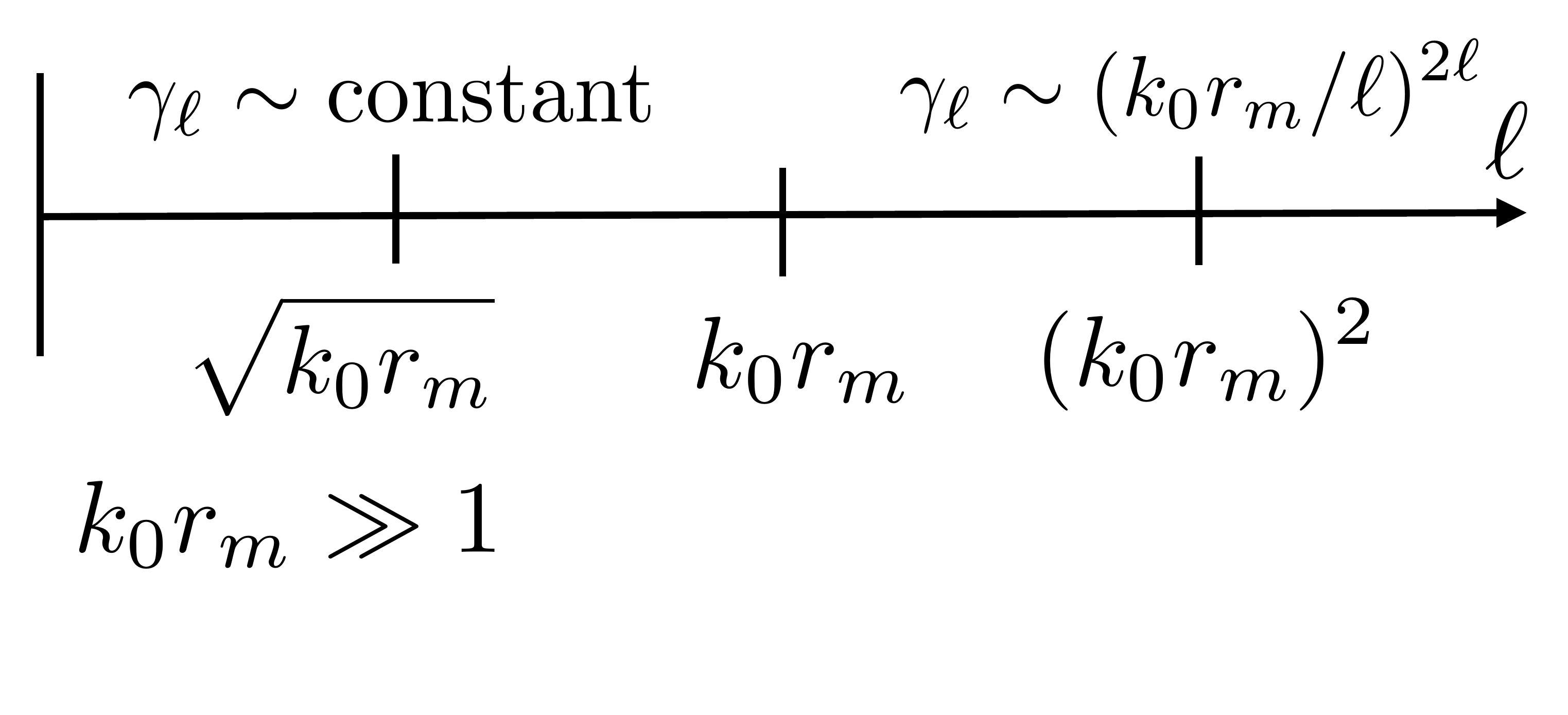}
\caption{Scaling of multipole emission rate $\gamma_\ell$ with increasing orbital angular momentum quantum number $\ell$ in the regime where the dipole approximation breaks down $k_0 r_m \gg 1$.}
\label{fig:s1}
\end{center}
\end{figure}

The spontaneous emission rate to emit light with orbital angular momentum $\ell$ during a radiative transition from $\ket{1,m}$ to $\ket{0,m+\ell}$ is given by Fermi's ``golden rule'' as
\be
\gamma_{\ell} = 2 \pi \sum_{k,k_\perp,i} |M_{m,m+\ell}^{\ell,k,i}|^2 \delta\Big(c \sqrt{k^2+k_\perp^2} - E_1 + E_0\Big)
\ee
The quantity $\gamma_\ell/\sum_\ell \gamma_\ell$ is plotted in Fig.~3(c) of the main text.  

To understand the scaling predicted by this equation we note that, in the generic case where $\ell_c \ll \lambdabar$ and $\ell \ll m,m'$, we can approximate the integral in Eq.~(\ref{eqn:m}) by replacing the photonic mode by its value at $r=r_m$.  This follows because the mode function $\bm{A}_{\ell,k}^i$  varies on the scale of $1/k_\perp > \lambdabar$, so it can be pulled out of the integral over the electronic wavefunctions, which are peaked at $r=r_m$ with a width given by $\ell_c$.  This implies the scaling
\be
|M_{m,m+\ell}^{\ell,k,i}|^2 \sim [J_\ell(k_\perp r_m)]^2.
\ee
As a result, we can find the scaling of $\gamma_\ell$ by looking at the different scalings of the Bessel function.  This is illustrated in Fig.~\ref{fig:s1} in the regime $k_0 r_m \gg1$.  

For $k_\perp r_m \ll \ell^2$, 
\be
|M_{m,m+\ell}^{\ell,k,i}|^2 \sim \cos^2(k_\perp r_m - \pi \ell /2-\pi/4),
\ee
 which oscillates with $\ell$.  However, in evaluating $\gamma_\ell$ we average over $k_\perp$, which washes out these oscillations.  As a result, in this regime $\gamma_\ell$ is approximately independent of $\ell$, in agreement  with the full calculations shown in Fig.~3(c) of the main text.   
In the opposite limit, $k_\perp r_m \gg \sqrt{\ell}$, 
\be
|M_{m,m+\ell}^{\ell,k,i}|^2 \sim \frac{(k_\perp r_m)^{2\ell}}{(\ell!)^2} \sim\bigg( \frac{ k_\perp r_m}{\ell}\bigg)^{2\ell},
\ee
where we used Stirlings approximation from above.
  In this regime, $\gamma_\ell$ recovers the typical behavior for higher-order multipole transitions  and decreases exponentially with $\ell$.


\bibliographystyle{../../apsrev-nourl}
\bibliography{../../GrCQED}

\begin{thebibliography}{61}
\bibitem[{foo()}]{footnote_ell}
\bibinfo{note}{Here the sum is restricted to positive $\ell$ because the
  transitions to negative $\ell$ are forbidden from Pauli blocking for the
  configuration in Fig.~1(c)}.
\end{thebibliography}

\end{document}